# Effect of projectile shape and interior structure on crater size in strength regime

T. Kadono[1 *], M. Arakawa[2], S. Tsujido[2], M. Yasui[2], S. Hasegawa[3], K. Kurosawa[4], K. Shirai[2], C. Okamoto[2], K. Ogawa[2,3], Y. Iijima[3], Y. Shimaki[3], K. Wada[4]

[1] Department of Basic Sciences, University of Occupational and Environmental Health, Kitakyusyu, Japan

[2] Department of Planetology, Kobe University, Kobe, Japan

[3] Institute of Space and Astronautical Science, Japan Aerospace Exploration Agency, Sagamihara, Japan

[4] Planetary Exploration Research Center, Chiba Institute of Technology, Narashino, Japan

*Correspondence: kadono@med.uoeh-u.ac.jp

**Abstract**

Experiments on crater formation in the strength regime were conducted using projectiles of various shapes with an aspect ratio of ~1, including both solid and hollow interiors. The surface diameter, inner (pit) diameter, and depth of the craters on basalt and porous gypsum targets were measured. Using the bulk density of the projectile, the surface diameter and depth for basalt and the pit diameter and depth for porous gypsum were scaled using the pi-scaling law for crater formation in the strength regime. The numerical code iSALE was used to simulate the impact of projectiles of various shapes and interior structure with similar bulk densities. Results show that the distributions of the maximum (peak) pressure experienced and particle velocity in the targets were similar regardless of projectile shape and interior structure, implying that the dimensions of the final craters were almost identical. This is consistent with the experimental results. Thus, we conclude that the size of the craters formed by the impact of projectiles with different shape and interior structure can be scaled using a conventional scaling law in the strength regime, using bulk density as projectile density.

**Keywords:** hypervelocity impact, crater formation, strength regime, hollow projectile, scaling law

## Introduction

An impact experiment was performed on the surface of the C-type asteroid 162173 Ryugu using an instrument called the Small Carry-on Impactor (SCI), carried by JAXA spacecraft Hayabusa2 (Arakawa et al., 2020). This SCI instrument launched a copper projectile with a mass of 2 kg, in the shape of a spherical shell - a hollow ball with a thickness of approximately 5 mm and a diameter of 13 cm. An important outcome is that the size of the artificial crater produced on Ryugu is well scaled by a conventional scaling law in the gravity regime when the bulk density is simply set as the projectile density in the scaling law (Arakawa et al., 2020), where the bulk density of the projectile 1.740 $g/cm^3$ is calculated by assuming that the projectile is a sphere with a diameter of 13 cm and a mass of 2 kg.

Projectiles with such complex structures including hollow interior have been used not only for Hayabusa2 but also for other planetary explorations in recent years, such as DEEP IMPACT (A'Hearn et al., 2005; Sugita et al., 2005; Kadono et al., 2007) and LCROSS (Colaprete et al., 2010; Schultz et al., 2010). However, to date, most impact experiments in laboratories have used solid projectiles, because the main purpose of these experiments has been simulating the impact of celestial bodies (e.g., Melosh, 1989). Hence, there have been a limited number of experiments in the field or in laboratories using projectiles with complex structures, such as the impact of hollow aluminum and nylon projectiles into sand and pumice targets in the context of clustered impacts (Schultz and Gault 1985) and the calibration experiments for the SCI impact (Saiki et al., 2017) and the LCROSS impact (Hermalyn et al., 2012). In particular, there have been few impact experiments using complex-structured projectiles and targets with strength. The SCI projectile in the Hayabusa2 mission collided with smaller boulders or grains on the surface of Ryugu, but larger boulders than the SCI projectile exist near the impact point and it could have possibly collided with these boulders. Hence, the impact of projectiles with hollow interiors to larger boulders with strength could occur in future missions, and we should investigate the craters caused by projectiles with hollow interiors on targets with strength. Thus, in this study, we conducted impact experiments using projectiles of cylindrical or spherical shapes with hollow or solid interiors and targets with strength, basalt and porous gypsum. In gravity regime, as shown in the impact of the SCI projectile, the crater efficiency of hollow projectiles can be scaled by the pi-scaling law for crater formation when the bulk density of the projectile is used (Schultz and Gault 1985). We investigated the sizes of the craters formed on the targets with strength to verify that, as in the gravity regime, crater size is scaled by a conventional scaling-law when bulk density is set as projectile density. Numerical simulation is done to support the experimental results.

## Experiment

We used four types of projectiles, as illustrated in Fig. 1: (a) a "closed" copper tube (one side is closed and the other is open), (b) an "open" copper tube (both sides are open), (c) a solid copper cylinder, and (d) a solid aluminum sphere. Tube-shaped projectiles were used because their thickness can be easily adjusted to match bulk density. The diameter, $D_p$, height, $H$, and bulk density, $\rho_p$, of the projectiles are listed in Table 1. The projectile was accelerated with a split-type sabot using a two-stage hydrogen-gas gun at Institute of Astronautical Science (ISAS) of JAXA (Kawai et al., 2010). Open tubes impacted on one side, and the solid cylinders impacted on one flat surface. For closed-tube projectiles, we conducted our experiments in two different ways: impact from the closed side or impact from the open side. Targets were porous gypsum and basalt, which have been extensively used to investigate the crater formation process on asteroid surfaces with high porosity (e.g., Yasui et al. 2012) and lunar and terrestrial surfaces covered with volcanic materials (e.g., Gault and Heitwit 1963), respectively. The tensile strength of these materials, $Y_t$, was 2.52 MPa (Yasui et al., 2012) and 19.3 MPa (Nakamura et al., 2007), respectively.

Target density, $\delta_t$, is listed in Table 1. Impact velocity, $V$, was approximately 2−4 km/s. Seventeen shots were successful. The impact conditions are listed in Table 1.

After the shot, the target surface was scanned by a high-resolution three-dimensional geometry-measurement system with a semi-conductor laser displacement sensor (COMS MAP-3D) at ISAS/JAXA. We obtained the topography of target surfaces every 0.2 mm interval. Based on this data, we determined the crater diameter and depth.

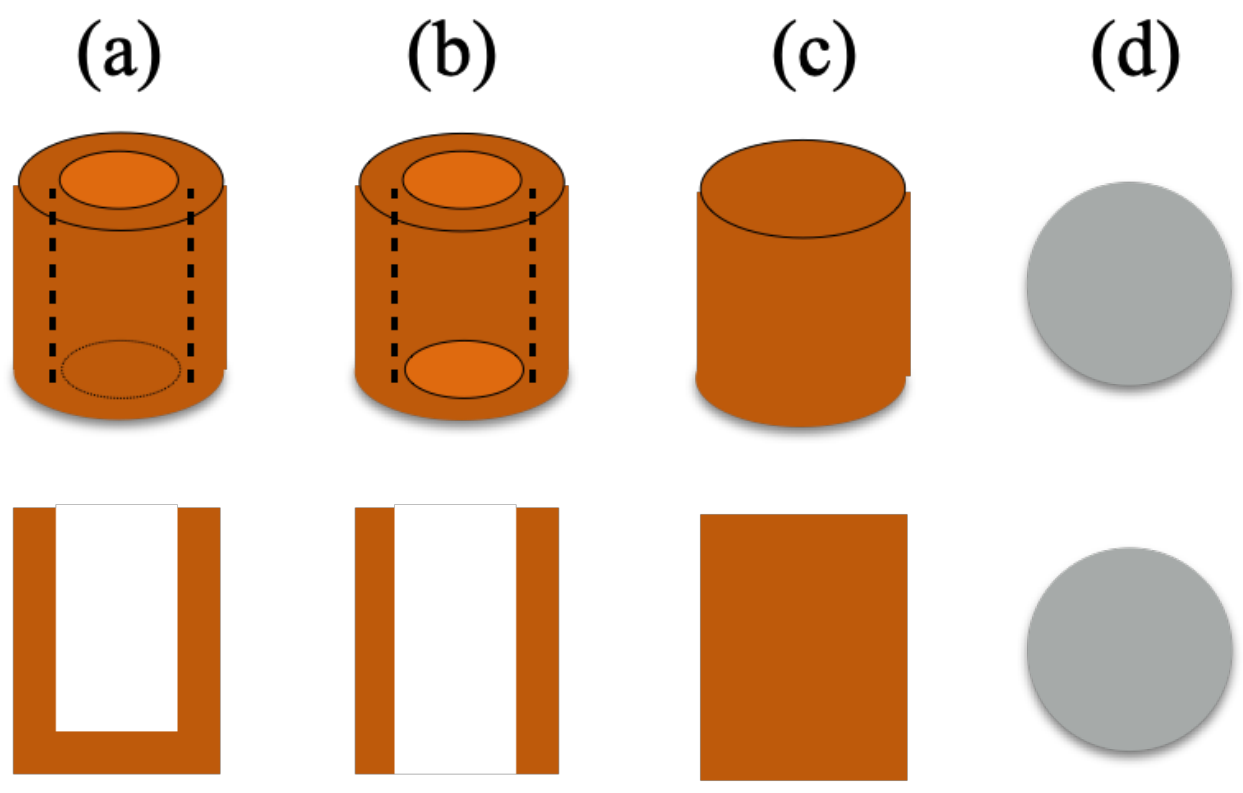

**Fig. 1** The shape of the projectiles used in our experiments: (a) a closed tube, (b) an open tube, (c) a solid cylinder, and (d) a solid sphere. The lower figure shows the cross section of each projectile. The projectiles are all fabricated from copper, except for the solid spheres, which are made of aluminum.

**Results**

Figure 2 shows typical crater profiles resulting from a closed copper tube impacting on porous gypsum (upper) and basalt (lower) targets. The craters on the porous gypsum targets consist of a deep pit with a spall region around the pit. We measured two kinds of diameters for the crater on the porous gypsum targets: the surface (spall) diameter $D_s$ at the surface of the targets and the inner (pit) diameter $D_i$, respectively, as indicated in the upper panel of Fig. 2. Shot numbers 0606-3 and 0606-4 were exceptions, as there was no clear spallation; therefore, only the diameter at the surface $D_s$ was measured and set $D_s = D_i$. On the other hand, there were no clear pits in the craters on the basalt targets. Hence, we measured only the surface diameter of the crater on the basalt targets. We also measured the depth, $d$, of the crater, which was measured from the surface to the deepest part of the crater.

Figure 3 shows crater diameters and depth normalized by projectile diameter, $D_p$, (a) $D_s/D_p$, (b) $D_i/D_p$, and (c) $d/D_p$, as a function of $\rho_p$. For comparison, previous data obtained using cylindrical solid nylon projectiles and basalt targets (Dohi et al., 2012) and a spherical solid nylon projectile and a porous gypsum target (Yasui et al., 2012) are also plotted. There appears to be no systematic difference depending on projectile type, projectile orientation, or whether the projectile structure (i.e., solid or hollow). The values, $D_s/D_p$, $D_i/D_p$, and $d/D_p$, are listed in Table 1.

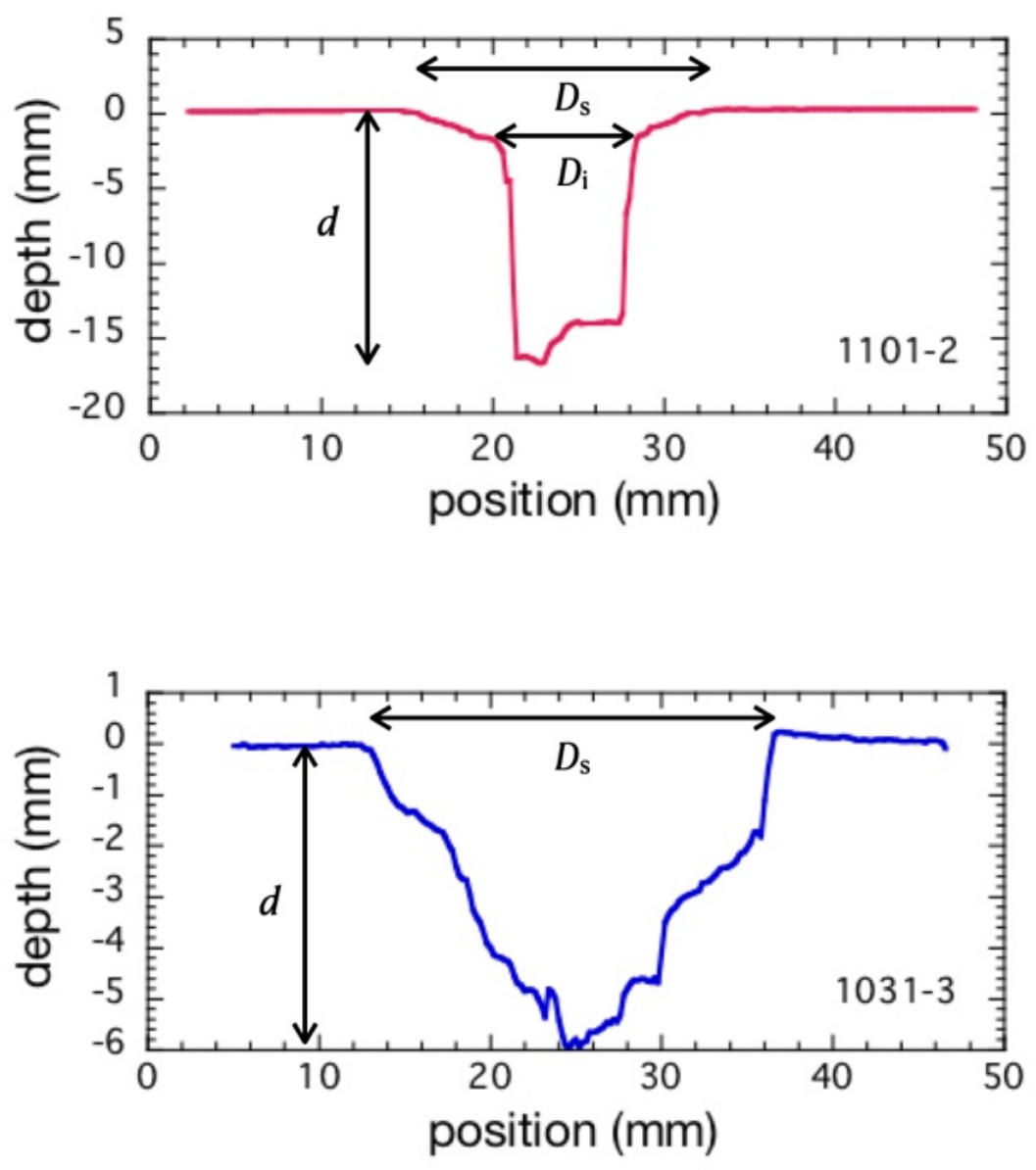

**Fig. 2** Typical crater profiles resulting from impact of a closed copper tube on (upper) porous gypsum target (shot # 1101-2) and (lower) basalt target (shot # 1031-3). The surface diameter, $D_s$, the inner (pit) diameter, $D_i$, and the depth, $d$, are defined as the diameter at the target surface, the diameter at the pit, and the distance from the surface to the deepest part of the crater, respectively. Note that the vertical scale is exaggerated.

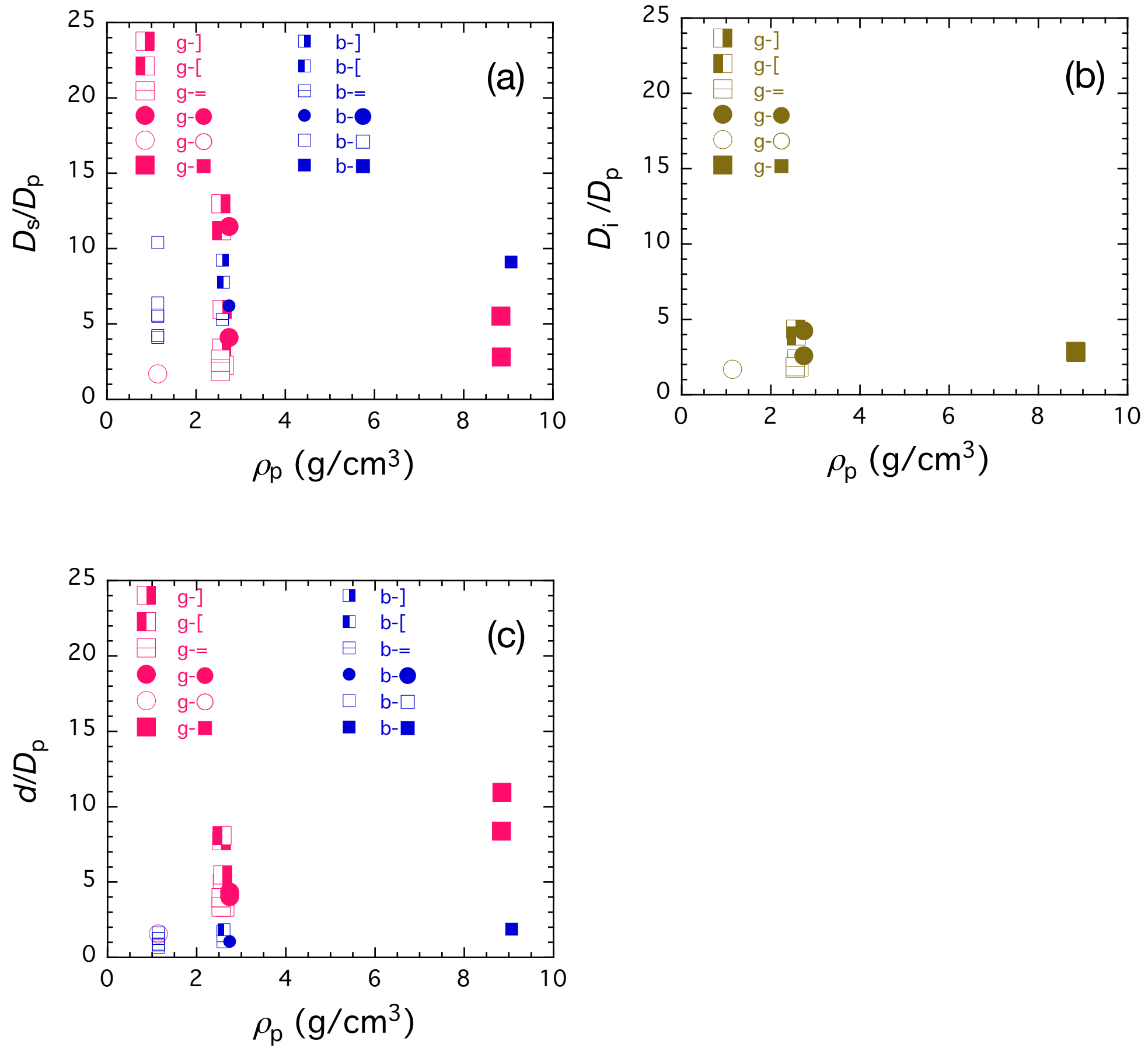

**Fig. 3** Crater diameters (surface and inner) and depth normalized by projectile diameter, $D_p$, as a function of projectile bulk density, $\rho_p$, in g/cm$^3$, (a) $D_s/D_p$, (b) $D_{pi}/D_p$, and (c) $d/D_p$. Symbols, "g-" and "b-", indicate porous gypsum and basalt targets, respectively. Projectile structures and orientations are also indicated; symbols "] ", "[", "=", "●", and "■" indicate the impact of closed tubes from the closed side, the impact of closed tubes from the open side, the impact of open tubes from one open side, the impact of solid spheres, and the impact of solid cylinders, respectively. Symbols "○" and "□" indicate the previous results of the impact of a spherical nylon projectile onto a porous gypsum block (1 point; Yasui et al. 2012) and of cylindrical nylon projectiles onto basalt blocks (6 points; Dohi et al. 2012).

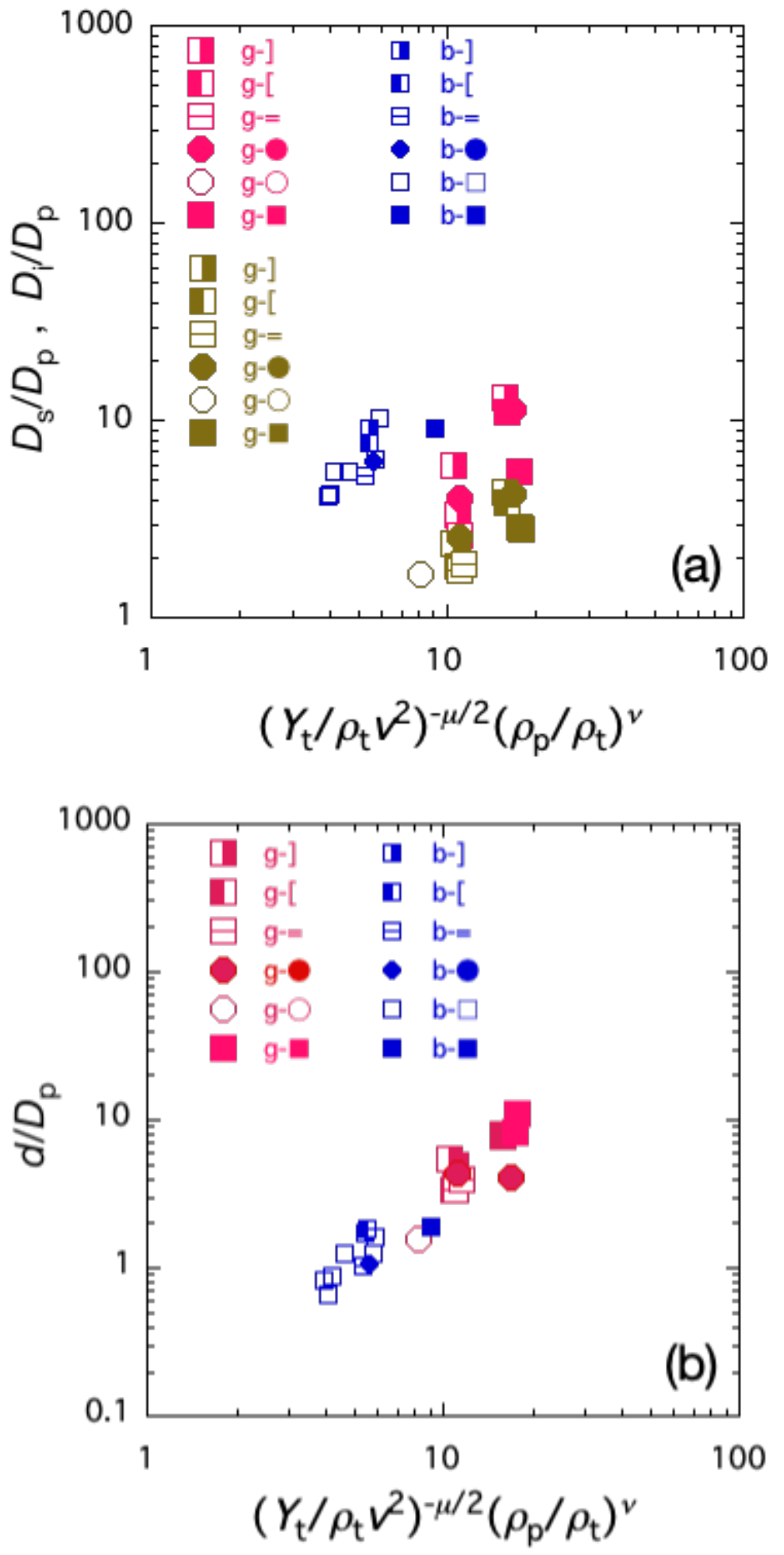

**Fig. 4** (a) Surface diameter, $D_s$, and inner diameter, $D_i$, normalized by $D_p$. (b) Crater depth, $d$, normalized by $D_p$. In both panels, the horizontal axis is the same, $\pi_3^{-\mu/2}\pi_4^{-\nu}$, where $\mu$ and $\nu$ are set to 0.55 and 0.4, respectively. The previous data of porous gypsum (Yasui et al. 2012) and basalt (Dohi et al. 2012) with nylon projectiles are also plotted. The symbols in the legend of each figure are as described for Fig. 3.

**Table 1**. Experimental conditions and results.

| Shot # | projectile | $\rho_p$ (g/cm$^3$) | $D_p$ (mm) | $H$ (mm) | target | $\delta_t$ (g/cm$^3$) | $V$ (km/s) | $D_s/D_p$ | $D_i/D_p$ | $d/D_p$ |
|---|---|---|---|---|---|---|---|---|---|---|
| 1030-1 | ] | 2.58 | 3.19 | 5.03 | G | 0.889 | 1.89 | 3.39 | 2.13 | 4.91 |
| 1030-2 | [ | 2.63 | 3.17 | 5.01 | G | 0.860 | 1.82 | 2.26 | 1.84 | 3.34 |
| 1030-3 | ] | 2.56 | 3.17 | 5.01 | G | 1.07 | 3.86 | 13.0 | 4.36 | 7.74 |
| 1031-1 | [ | 2.57 | 3.19 | 5.02 | G | 1.07 | 3.88 | 11.2 | 3.90 | 8.08 |
| 1031-2 | ] | 2.59 | 3.18 | 5.01 | B | 2.71 | 1.91 | 9.24 | - | 1.73 |
| 1031-3 | [ | 2.62 | 3.18 | 5.01 | B | 2.73 | 1.90 | 7.77 | - | 1.84 |
| 1031-4 | S | 2.74 | 3.2 | 3.2 | G | 0.940 | 1.87 | 4.10 | 2.58 | 4.34 |
| 1031-5 | S | 2.74 | 3.2 | 3.2 | B | 2.73 | 1.92 | 6.21 | - | 1.06 |
| 1031-6 | S | 2.74 | 3.2 | 3.2 | G | 1.14 | 4.19 | 11.5 | 4.24 | 4.03 |
| 1101-1 | C | 8.84 | 3.0 | 1.65 | G | 0.880 | 1.79 | 5.53 | 2.89 | 8.38 |
| 1101-2 | ] | 2.58 | 3.18 | 5.02 | G | 1.14 | 1.84 | 5.97 | 2.38 | 5.48 |
| 1101-3 | ] | 2.58 | 3.18 | 5.02 | G | 0.843 | 4.05 | 3.39 | 2.13 | 4.91 |
| 0606-2 | = | 2.55 | 3.0 | 2.0 | G | 1.12 | 2.02 | 2.66 | 1.76 | 3.34 |
| 0606-3 | = | 2.55 | 3.0 | 2.0 | G | 0.936 | 2.09 | 1.86 | 1.86* | 3.97 |
| 0606-4 | C | 8.84 | 2.0 | 1.3 | G | 0.928 | 1.90 | 2.81 | 2.81* | 11.0 |
| 0925-1 | = | 2.59 | 3.0 | 2.0 | B | 2.67 | 1.82 | 5.30 | - | 1.04 |
| 0925-2 | C | 9.06 | 2.0 | 1.1 | B | 2.72 | 1.93 | 9.12 | - | 1.88 |

Projectile:
]: copper closed tube impacting at the closed side
[: copper closed tube impacting at the open side
=: copper open tube
S: aluminum solid sphere
C: copper solid cylinder
Target:
G: porous gypsum
B: basalt
*No spallation and we set $D_s = D_i$

## Discussion

### Scaling by conventional crater scaling law

The pi-scaling law for crater size in the strength regime describes crater size, $R$, using the dimensionless parameters $\pi_R = R(\delta_t/m)^{1/3}$, $\pi_3 = (Y_t/\delta_t V^2)$, and $\pi_4 = (\delta_t/\delta_p)$ as

$$\pi_R = k_1 \pi_3{}^{-\mu/2} \pi_4{}^{(1-3\nu)/3}\ , (1)$$

where $\delta_t$, $m$, $V$, and $\delta_p$ are target density, projectile mass, impact velocity, and projectile density, respectively, and $k_1$, $\mu$, and $\nu$ are empirical parameters related to the point source approximation and determined by laboratory experiments (Housen and Holsapple, 2011). Since $\pi_R \sim (R/D_p)\pi_4{}^{1/3}$, where $D_p$ is the projectile diameter, Eq. (1) becomes

$$\left(\frac{R}{D_p}\right) \sim \left(\frac{\delta_p}{\delta_t}\right)^{\nu} \left(\frac{Y_t}{\delta_t V^2}\right)^{-\mu/2}\ . (2)$$

Figure 4 shows normalized crater size as a function of the right-hand side of Eq. (2). We set the empirical parameters, $\mu$ and $\nu$, to typical values previously obtained for hard rocks of 0.55 and 0.4, respectively (Holsapple, 1993; Housen and Holsapple, 2011), and the projectile density $\delta_p$ to the bulk density of projectiles, $\rho_p$. Previous data are also plotted for basalt (Dohi et al., 2012) and porous gypsum (Yasui et al., 2012). For basalt targets, the data of surface diameter $D_s/D_p$ in (a) and depth $d/D_p$ in (b) are in good agreement with the previous data and show a linear relationship. For porous gypsum, the data for the surface (spall) diameter $D_s/D_p$ is scattered, but the data for inner (pit) diameter $D_i/D_p$ and depth $d/D_p$ appear to be linearly correlated and in good agreement with the previous data using different types of projectiles. Thus, even when projectiles have varied shapes, crater sizes in the strength regime can be scaled using a conventional scaling law previously established, using bulk density as $\rho_p$. Note that spallation in porous gypsum targets often produces large fragments (e.g., Fig. 3 in Suzuki et al. 2018). Such a process with large crack propagation is highly probabilistic; hence, the surface (spall) diameter $D_s/D_p$ for porous gypsum may scatter. This may be the reason why spallation occurred in shot number 0606-2 but not in 0606-3, even though the impact conditions for these shots were almost the same. On the other hand, the reason of no spallation in shot number 0606-4 would be different. The density ratio of projectile to target in shot number 0606-4 was very high ~9: such high density-ratio generally causes a carrot shaped crater without spallation (e.g., Love et al. 1993; Kadono et al. 2012)). Furthermore, target differences cannot be scaled by only strength and density. Other parameters should be considered, but these are beyond the scope of this paper and are not discussed further.

As an application of our results, we consider the case if the SCI projectile collides with a boulder on Ryugu (e.g., "SB" boulder with a size of 5 m existing in the vicinity of the SCI impact point as shown in Arakawa et al. 2020). The left-hand side of Eq. (2) becomes ~6.6, when setting $\rho_p$ and $V$ to 1.74 g/cm$^3$ and 2 km/s for the SCI projectile (Arakawa et al. 2020) and $\rho_t$ and $Y_t$ to 1 g/cm$^3$ and 1.7 MPa for the SB boulder (Kadono et al. 2020). In this case, if porous gypsum can simulate the boulders with high porosity on Ryugu, Fig. 4 indicates that $D_i/D_p$ and $d/D_p$ are ~1, respectively, suggesting that the crater is much smaller than the actual crater size that formed in the gravity regime, and would have been extremely difficult to find.

**Numerical simulations**

We investigated the crater formation process under the conditions experimented in this study using a general-purpose shock-physics code, iSALE-2D (Wünnemann et al., 2006) to confirm the limited dependence of crater size on projectile shape. This code is an extension of the SALE code (Amsden et al., 1980), which is capable of modelling shock processes in geological materials (Melosh et al., 1992; Ivanov et al., 1997; Collins et al., 2004; Wünnemann et al., 2006).

Three types of projectiles were considered: closed copper tubes, open copper tubes, and solid aluminum spheres, corresponding to projectiles described in Figs. 1a, 1b, and 1d, respectively. The diameter and height of all the projectiles were set to 3.2 mm. To simulate the impact of projectiles having a similar bulk density, the thicknesses of the closed and open tubes were set to 0.20 and 0.25 mm, respectively, resulting in the bulk densities of the closed and open tubes being 2.53 and 2.58 g/cm$^3$, respectively.

Four types of impacts were simulated: (a) a closed tube impacting on the closed side, (b) a closed tube impacting on the open side, (c) an open tube on one open side, and (d) a solid aluminum sphere. These projectiles collided perpendicularly on basaltic target flat surfaces along the central axis. Impact velocity was set at 2 km/s.

The calculation settings are summarized as follows: we used the two-dimensional version of iSALE, which is referred to as iSALE-Dellen (Collins et al., 2016) to simulate the vertical impacts performed in the experiments. We used the Tillotson equations of state (EOS) for copper and aluminum (Tillotson, 1962) and the "Analytical" EOS (ANEOS) (Thompson and Lauson, 1972) for basalt (Pierazzo et al., 2005; Sato et al., 2021). We also employed a constitutive model to calculate the elastoplastic behaviors of both shocked projectiles and targets. The Johnson-Cook model (Johnson and Cook, 1983) was used for metal projectiles. We used the "ROCK" model implemented in the iSALE for the basalt target, which is a combination of the Drucker-Prager model (Drucker and Prager, 1952) for damaged rocks and the Lundborg model for intact rocks (Lundborg, 1968). The two models were coupled with a damage parameter ranging from 0 to 1, depending on the total plastic strain (e.g., Ivanov et al., 1997; Collins et al., 2004). It was not feasible for the numerical integrations to continue until the end of crater formation and hence, we addressed the peak pressure and resultant particle-velocity distributions at a given time. It has been shown that iSALE represents the maximum (peak) pressures experienced at each position in the targets caused by the shock wave well (Nagaki et al., 2016). The compression to a sufficient pressure by the shock wave and the release from the pressure by the rarefaction wave cause the fragmentation of the target material. Crater depth would correspond to a position experienced by a critical peak pressure value. On the other hand, crater diameter is directly related to fragmentation at the target surface, caused by a tensile phase due to the rarefaction wave (e.g. Melosh, 1989). Since the rarefaction wave leads to upward motion of the target materials, the distribution of particle velocity near the surface should represent the extent of fragmentation near the surface. If the peak pressure and particle velocity distributions do not strongly depend on projectile shape and interior, the dimensions of the final crater are expected to be similar. To accurately reproduce the dependence of peak pressure distribution on projectile shape and interior in the simulation, we divided the wall thickness into 20 cells, resulting in >160 cells per projectile radius for the three copper tubes. We inserted Lagrangian tracer particles into the computational cells and stored the maximum pressures experienced at a given time and temporal particle velocity in the simulations. The input parameters of the material models and calculation settings are summarized in Tables 2–4.

**Table 2**. Input parameters of the material models for the metal projectiles. Detailed descriptions of the parameters can be found in the iSALE manual (Collins et al., 2016).

| | Copper | Aluminum |
|---|---|---|
| EOS model | Tillotson EOS[a] | Tillotson EOS[a] |
| Poisson's ratio | 0.34[b] | 0.33[c] |
| Melting temperature (K) | 1,356[d] | 978[e] |
| Simon parameter *a* (GPa) | 49.228[d] | 7.98[e] |
| Simon parameter *c* | 1.027[d] | 0.57[e] |
| Specific heat (J $K^{-1}$ $kg^{-1}$) | 392.5[f] | 896[f] |
| Johnson-Cook parameter *A* (MPa) | 90[g] | 244[h] |
| Johnson-Cook parameter *B* (MPa) | 292[g] | 488[h] |
| Johnson-Cook parameter *N* | 0.31[g] | 0.5[h] |
| Johnson-Cook parameter *C* | 0.025[g] | 0.0[h] |
| Johnson-Cook parameter *m* | 1.09[g] | 3[h] |
| Reference temperature (K) | 293 | 293 |

a. Tillotson (1962)
b. Köster and Franz (1961)
c. We used an aluminum alloy. The Poisson ratio was taken from the data compiled on the web page (http://asm.matweb.com/search/GetReference.asp?bassnum=ma6061t6) provided by ASM Aerospace Speciation Metals Inc. The database was constructed based on the information provided by the Aluminum Association Inc.
d. The pressure-dependent melting temperature of copper presented in Japel et al. (2005) was fitted by the Simon equation (e.g., Poirier, 1991; Wünnemann et al., 2008).
e. Hänström and Lazor (2000).
f. The Dulong-Petit values were used.
g. Johnson and Cook (1983)
h. Pierazzo et al. (2008)

**Table 3**. Material model parameters for basalt targets.

| | Basalt* |
|---|---|
| EOS model | ANEOS |
| Poisson's ratio | 2.5 |
| Melting temperature (K) | 1,360 |
| Simon parameter *a* (GPa) | 4.5 |
| Simon parameter *c* | 3.0 |
| Thermal softening coefficient | 0.7 |
| Cohesion (intact) (MPa) | 20 |
| Frictional constant (intact) | 1.4 |
| Cohesion (damaged) (MPa) | 0.01 |
| Frictional constant (damaged) | 0.6 |
| Limiting strength (GPa) | 2.5 |
| Minimum failure strain | $10^{-4}$ |
| Constant for the damage model | $10^{-11}$ |
| Threshold pressure for the damage model (MPa) | 300 |

* The parameters used in this study are the same as used in Bowling et al. (2020).

**Table 4**. Calculation settings.

| Parameters and settings * | Values |
|---|---|
| Computational geometry | Cylindrical coordinate |
| Number of computational cells in the *R* direction | 2000 |
| Number of computational cells in the *Z* direction | 3000 |
| Number of cells for the extension in the *R* direction | 400 |
| Number of cells for the extension in the *Z* direction | 400 |
| Extension factor | 1.02 |
| Grid spacing (m $grid^{-1}$) | $10^{-5}$ |
| Artificial viscosity $a_1$ | 0.24 |
| Artificial viscosity $a_2$ | 1.2 |
| Impact velocity (km $s^{-1}$) | 2.0 |
| High-speed cutoff | 2-fold impact velocity |
| Low-density cutoff (kg $m^{-3}$) | 100 |

*Detailed descriptions of the parameters can be found in the iSALE manual (Collins et al., 2016).

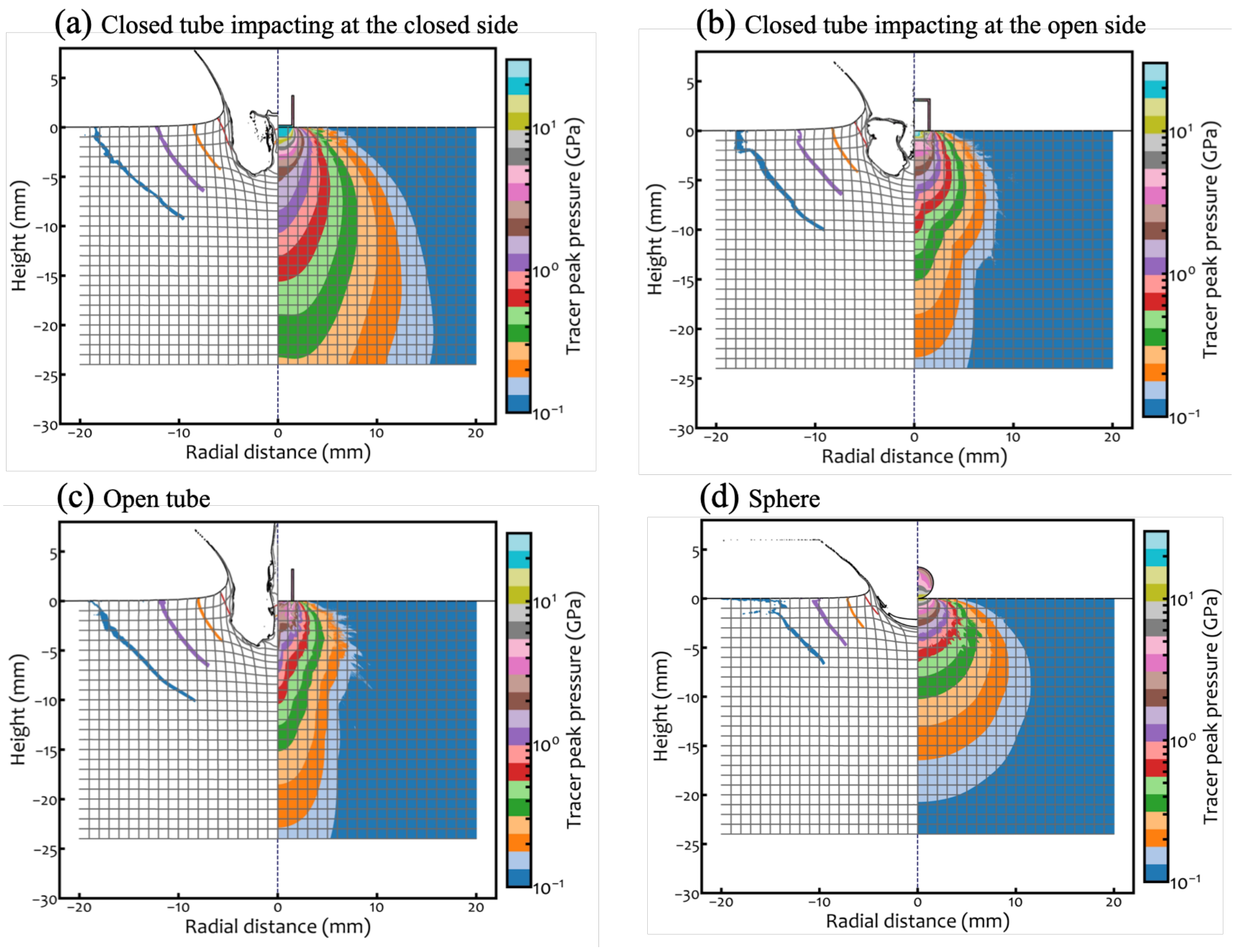

**Fig. 5** Results of the numerical simulations for (a) a closed copper tube impacting at the closed side, (b) a closed copper tube impacting at the open side, (c) an open copper tube, and (d) a solid aluminum sphere. The projectile collides perpendicularly at the origin of the coordinate from the top. The right half shows the contour of the maximum (peak) pressures experienced. The left half shows the contour of the upward particle velocity; the blue, purple, orange, and red lines indicate 3, 10, 30, and 100 m/s, respectively.

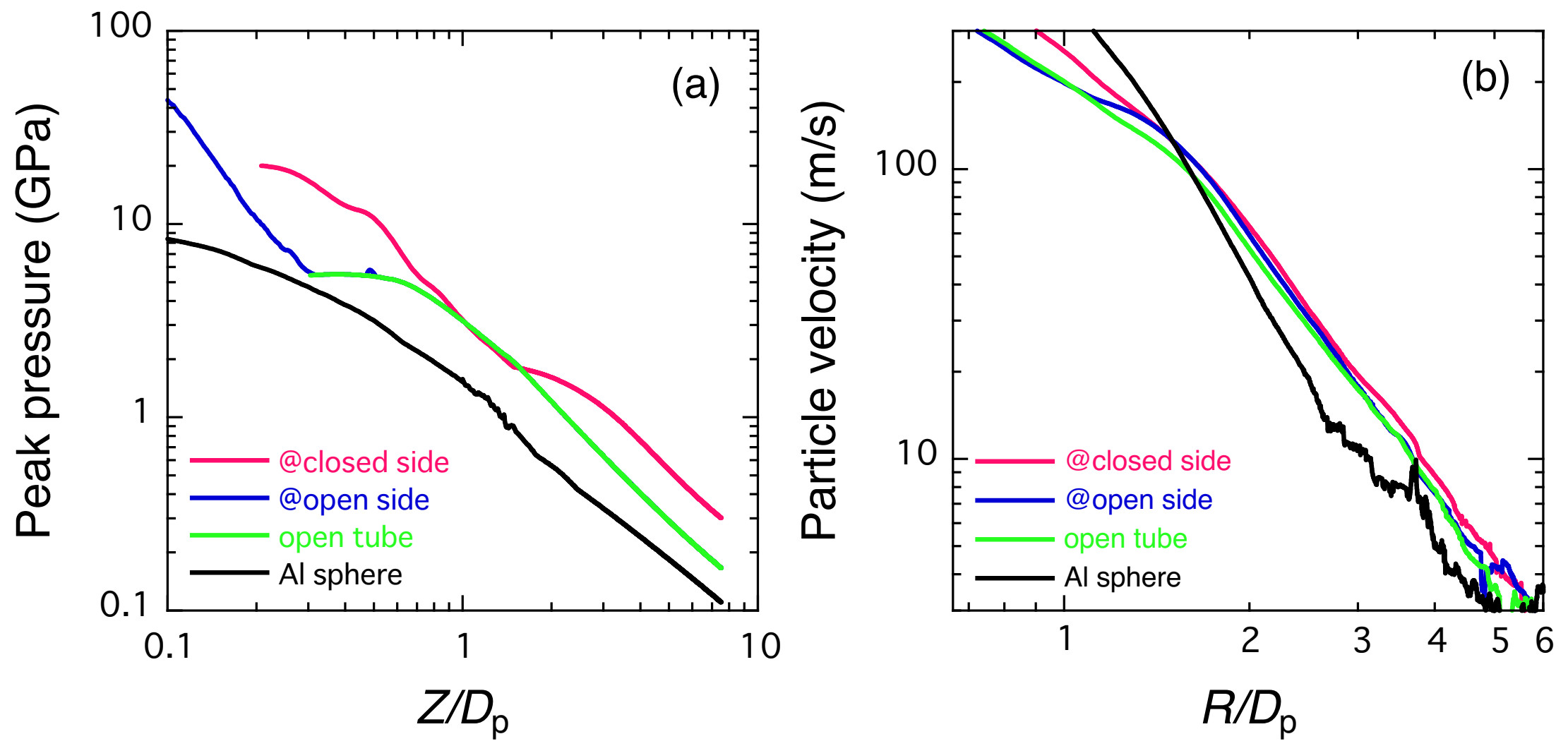

**Fig. 6** Profiles of (a) peak pressure and (b) particle velocity along the central axis and the target surface, respectively, for the numerical results shown in Fig. 5. The profiles for the copper closed-tube projectile impacting at the open side and the open tube projectile are very similar and hence, the lines are almost overlapping.

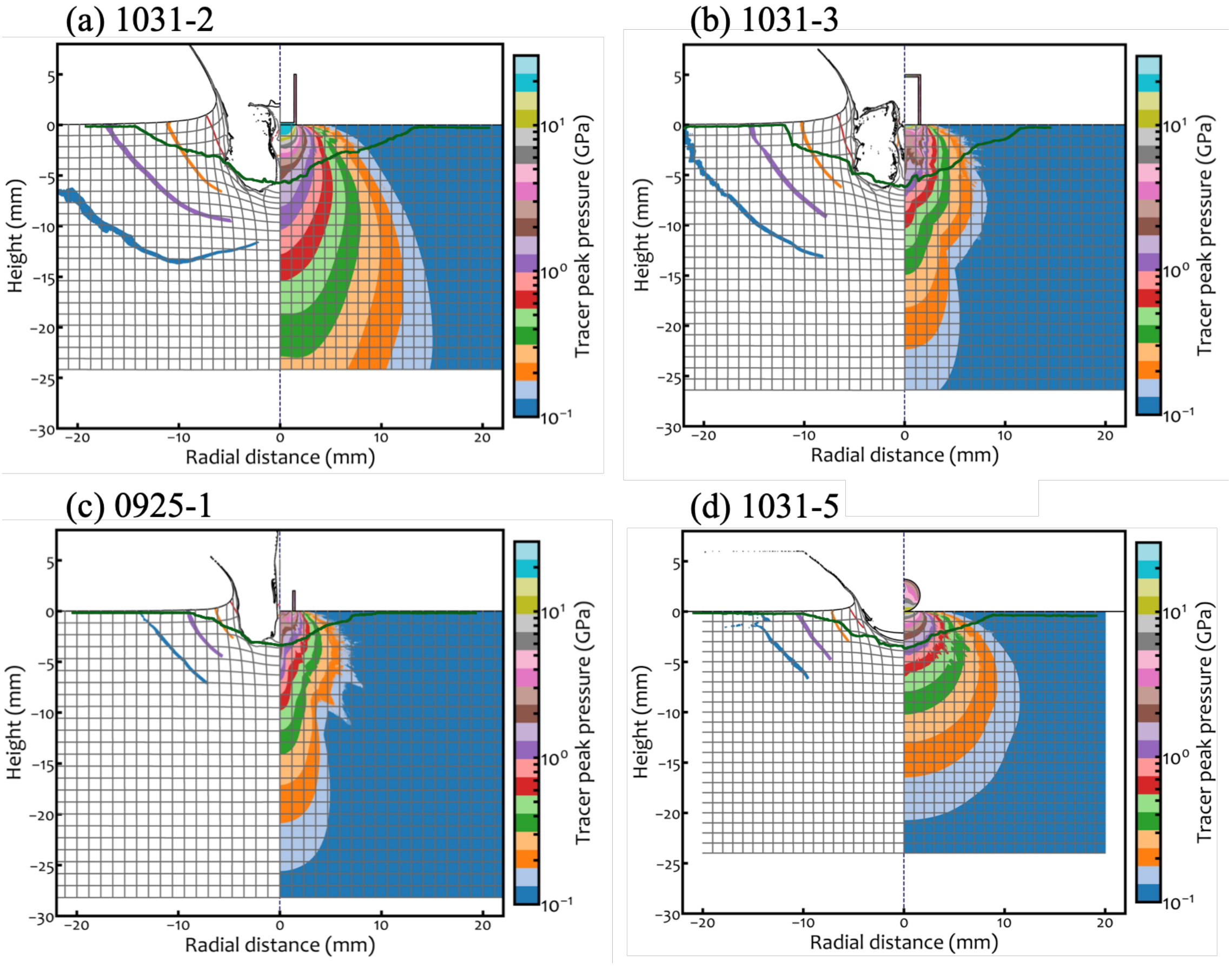

**Fig. 7** Results of the numerical simulations corresponding the experiments of (a) a closed copper tube impacting at the closed side (1031-2), (b) a closed copper tube impacting at the open side (1031-3), (c) an open copper tube (0925-1), and (d) a solid aluminum sphere (1031-5). The right half shows the contour of the maximum (peak) pressures experienced. The left half shows the contour of the upward particle velocity; the blue, purple, orange, and red lines indicate 3, 10, 30, and 100 m/s, respectively. The crater profile in the experiment with the corresponding impact condition is shown for comparison (green curve).

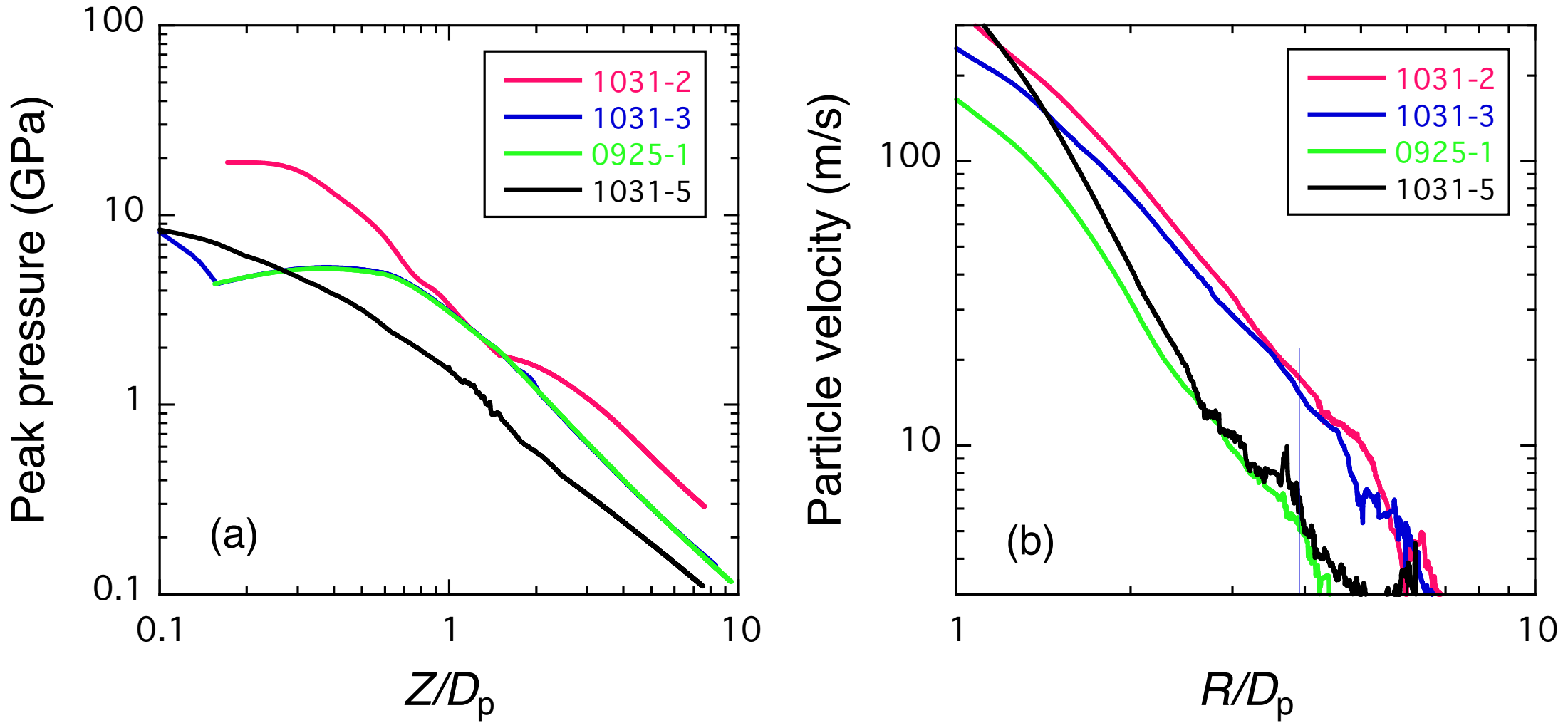

**Fig. 8** Profiles of (a) peak pressure and (b) particle velocity along the central axis and the target surface, respectively, for the results shown in Fig. 7. The peak pressure profiles for 1031-3 (the closed tube impact at open side) and 0925-1 (open tube) are similar and hence almost overlap. The experimental results of crater depth and crater radius are indicated as vertical lines (the line and profile with the same color are corresponding).

Figure 5 shows the results of the calculations: contours of the peak pressure experienced (right) and particle velocity (left). Note that we only used tracers with upward velocities in this plot (hereafter referred to as upward particle velocity). It appears that in any case, pressure decreases along the central axis in the same way, and that the contour lines of the upward particle velocity are distributed at a similar location. To evaluate the peak pressure and particle velocity more quantitatively, we obtained the profiles of the peak pressure and particle velocity along the central axis and target surface, respectively (the analyses are described in the section of S1 in Supplementary Materials (SM) in detail). Figure 6 shows that the profiles of (a) peak pressure experienced and (b) particle velocity for the results shown in Fig. 5 along the central axis ($Z$-axis) and the target surface ($R$-axis), respectively (strictly a line 5 cells away from the $Z$ and $R$ axes; see the section S1 in SM). These show that the profile for each impact condition decreases with a similar slope at distances greater than projectile diameter $D_p$ and that the difference between the profiles is within a factor of ~2−3. This implies that when projectile bulk density and impact velocity are the same, the pressure of shock wave detached from isobaric core is almost independent of the projectile shape and internal structure. Even if projectile has an internal structure (e.g., voids), since shock pressure is much higher than the strength of projectile, its shape and the internal structure are completely crushed, and the compressed density and average shock pressure are independent of initial shape and internal structure. Moreover, attenuation of detached shock waves depends on the geometrical expansion of shock waves and elastic-plastic properties of target materials. Therefore, shock pressure and attenuation are independent of initial projectile structure. Thus, even if projectile shape and internal structure vary, the crater depth and diameter become almost the same when the bulk density and diameter of projectile and impact velocity are the same. Note that the profiles also show that the result for the aluminum solid projectile is similar to those for copper projectiles with internal structure. This suggests that even when projectiles are made of different materials, shock impedance becomes similar if bulk density is the same. More systematic investigation is necessary to understand why similar pressures are generated by porous projectiles with internal structure and solid projectiles of different materials when bulk density is the same.

We also simulated the impacts corresponding to the experiments with shot number 1031-2 (closed copper tube colliding at the closed end), 1031-3 (closed copper tube colliding at the open end), 0925-1 (open copper tube), and 1031-5 (solid aluminum sphere). The results are shown in Figs. 7a–7d. The contours of the experienced peak pressure and particle velocity are shown on the right and left halves, respectively. For comparison, the crater profile in the corresponding experiment is overlaid (green curve).

The shape of the cavity in the targets is very different from that of the final crater and does not become similar thereafter, although the size is comparable. It seems that it is still difficult to reproduce the exact shape of a crater in targets with strength. Figure 8 shows the profiles of (a) peak pressure experienced and (b) particle velocity for the results shown in Fig. 7 along the central axis and target surface, respectively (strictly a line 5 cells away). The crater depth and radius normalized by $D_p$ obtained in the experiments are also indicated. The peak pressure and particle velocity corresponding to the experimental results of crater depth and radius are shown in Fig. 9. The depth and radius of the final crater corresponded to ~2 GPa and ~10 m/s, respectively, in each case. In this figure, the results of the shot number 0925-2 and one of the shots in Dohi et al. (2012) (090528-1) are also plotted (Note that the calculations for 0925-2 and 090528-1 in iSALE were conducted with low resolution and that nylon projectile was used in 090528-1. These calculations in iSALE with low resolution and nylon are described in the sections of S2 and S3 in SM, respectively). Even though the projectile density is different, the corresponding pressure is similar, and so is the corresponding particle velocity. The averages of the corresponding peak pressure and particle velocity are 1.9±0.6 GPa and 13.0±3.6 m/s, respectively. The compressive strength of basalt has been measured to be several hundred megapascals (e.g., Lockner, 1995), which is slightly lower than the pressure corresponding to the crater depth. However, the compressive strength values were obtained using static compression tests, and the effect of strain rate could explain this difference (e.g., Kimberley et al., 2013). Therefore, the maximum depth of the final craters in the strength-dominated regime was determined by the balance between the intensity of the compressive pulse and the compressive strength of the target materials. On the other hand, the upward particle velocity $u_p$ of ~10 m/s corresponds to the tensile stress, $T$, of several tens of megapascals, which is estimated from $\delta_t C_0 u_p$, where $\delta_t$ and $C_0$ are the density and bulk sound velocity of basalt, 2.71 g/cm$^3$ (the average value for the targets in our experiments) and 2.60 km/s (e.g., Melosh, 1989), respectively. This is similar to the tensile strength of basalt (e.g., Lockner, 1995), suggesting that crater diameter is related to the tensile process. The results, namely that crater depth and diameter are related to the compressive and tensile strengths of the targets, are consistent with our current understanding of the cratering process (e.g., Melosh, 1989).

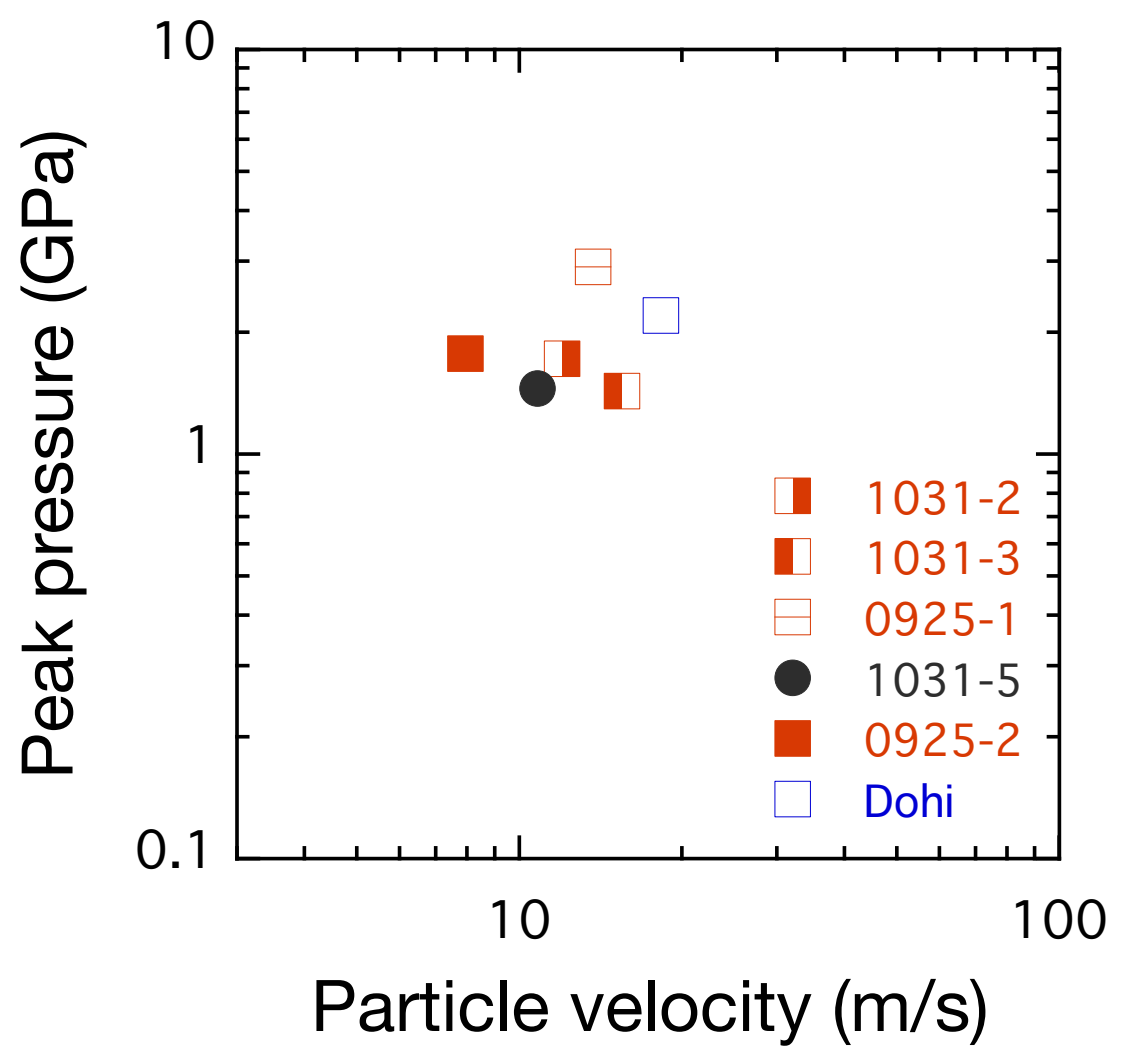

**Fig. 9** Peak pressure and particle velocity corresponding to the experimental results of crater depth and radius. The results of 0925-2 and one of the shots in Dohi et al. (2012) (090528-1) are also plotted.

**Summary**

Crater formation experiments were conducted using projectiles of various shapes and basalt and porous gypsum targets. The surface (spall) diameter, inner (pit) diameter, and depth of the craters were measured. The surface diameter and depth for basalt and the pit diameter and depth for porous gypsum were scaled using pi-scaling law for crater formation in the strength regime, when using the bulk density of projectiles. Consequently, the result is the same as in the gravity regime. The numerical code iSALE was used to simulate the impact of projectiles of various shapes with similar bulk densities. The contours of the maximum pressure experienced and the particle velocity in the targets obtained by the calculations were similar regardless of the projectile shape and interior. This implies that the dimensions of final craters were almost identical and is consistent with the experimental results. Thus, we conclude that the size of the crater formed by the impact of projectiles with different shape and interior structure can be scaled using a conventional scaling law in the strength regime, when the bulk density is set as the projectile density. Note that despite varied projectile shape and interior structure, the aspect ratio of projectiles used in this study was approximately 1. If the aspect ratio is larger, the conventional laws may not be able to explain crater sizes.

When new exploration methods such as SCI-type impact experiments are to be used in future missions to investigate the properties of the object being explored (e.g., the strength of boulders), the results of our study should provide constraints on the design of an impactor.

**Acknowledgments**

The authors are grateful to R. Honda, N. Sakatani, T. Saiki, Y. Imamura, H. Yano, S. Nakazawa, N. Hirata, and Y. Takagi for their support as a member of the SCI team of the Hayabusa2 mission in the early stages of this research and A. I. Suzuki and M. Tabata for supporting the impact experiments. We thank the developers of iSALE, including G. Collins, K. Wünnemann, B. Ivanov, J. Melosh, and D. Elbeshausen. We also thank Tom Davison for the development of pySALEPlot. The shock physics modelling was in part carried out on PC cluster at Center for Computational Astrophysics, National Astronomical Observatory of Japan. We also thank two anonymous reviews for useful comments. This work was supported by ISAS/JAXA as a collaborative program with the Hypervelocity Impact Facility.

**Authors' contributions**

TK, MA, ST, MY, SH, KS, CO, KO, YI, and KW conducted the impact experiments and contributed to data preparation, interpretation of results, and writing of the manuscript. KK and YS performed the numerical simulations, and drafted the initial manuscript. TK designed the paper and completed the manuscript. All authors read and approved the final manuscript.

**Funding**

This study was not supported.

**Availability of data and materials**

The datasets used and/or analyzed during the current study are available from the corresponding author on reasonable request.

**Declarations Competing interests**

The authors declare that they have no competing interests.

Effect of projectile shape and interior structure on crater size in strength regime

T. Kadono, M. Arakawa, S. Tsujido, M. Yasui, S. Hasegawa, K. Kurosawa, K. Shirai,

C. Okamoto, K. Ogawa, Y. Iijima, Y. Shimaki, K. Wada

# Supplementary Materials

**S1. Peak pressure and particle velocity**

We conducted post analyses of the iSALE output to obtain the peak pressure and particle velocity distributions (Figs. 6 and 8). These profiles after certain times after the impacts with tracer probes were quantitatively obtained with the pySALEPlot. We obtained the peak pressure distributions along the line parallel to the central axis away from 5 cells as a function of the initial height. The non-zero distance was chosen for the vertical measurement line because the variables in the region too close to the central axis in cylindrical coordinate is not reliable. The peak particle velocity distribution as a function of the initial radial distance was also obtained along the horizontal measurement line parallel to the target surface. Since top a few cells from the target surface suffer the effects of smearing due to artificial viscosity used in the iSALE [e.g., Kurosawa et al. 2018], we used a tracer probe located 5 cells below the target surface. We stored the maximum (peak) pressure and particle velocities in both radial and vertical directions on tracers during the simulations.

**S2. Low resolution numerical calculation for the shot number 0925-2 and one of the shots by Dohi et al. (2012) (090528-1)**

We conducted two low-resolution numerical simulations to model vertical impacts of solid projectiles (corresponding to the experimental run numbers 0925-2 and 090528-1). We used solid cylinders made of copper (0925-2) and Nylon 66 (090528-1) and basalt blocks. As discussed in the main text, the shortest side in the projectile dimensions was divided into 20 cells to accurately solve the peak pressure distribution during impacts in this study. If we used solid projectiles with the aspect ratio of ~1, the required grid size is much larger than that employed in the simulations supporting Figs. 5 and 7 shown in the main text. The input parameters of the material model pertaining to copper are listed in Table 2 in the main text. We used the Tillotson EOS with the parameter set pertaining to Nylon 66 for the projectile established in the section S3. Since material models for polymers have not been implemented into the stable release of the iSALE shock physics code, we conducted two different runs with and without a constant yield strength of 100 MPa for the Nylon 66 projectile [Haynes, 2010] used in #090528-1. We confirmed that the constant yield strength in the Nylon 66 projectile does not largely change impact outcomes. Thus, we only show the results in the case without any strength in the projectile. The calculation settings are summarized in Tables S1.

**Tables S1.** Calculation settings. Note that the settings are basically same as listed in Table 4 in the main text. Here, we only show the differences from the high-resolution models.

| Experimental run ID | 0925-2 | 090528-1 |
|---|---|---|
| Projectile materials | Copper | Nylon |
| Projectile diameter (mm) | 2.0 | 6.35 |
| Projectile length (mm) | 1.1 | 5.0 |
| Number of computational cells in the *R* direction | 750 | 750 |
| Number of computational cells in the *Z* direction | 800 | 800 |
| Number of cells for the extension in the *R* direction | 200 | 200 |
| Number of cells for the extension in the *Z* direction | 200 | 200 |
| Grid spacing (m grid$^{-1}$) | $2.75 \times 10^{-5}$ | $1.25 \times 10^{-4}$ |
| Impact velocity (km s$^{-1}$) | 1.93 | 1.98 |

**S3. Tillotson equations of state for Nylon 66**

We used a cylinder made of Nylon 66 as a projectile in the experimental run #090528-1. We constructed an equations of state (EOS) model pertaining to Nylon 66 to model this experiment in shock physics modeling. The Tillotson EOS format [Tillotson, 1962] was chosen because it allows to address a wide region in a *P*–*ρ*–*E* space, where *P* is pressure, *ρ* is density, and *E* is internal energy. Pressure is described as a function of *ρ* and *E* follows,

$$P(\rho, E) = \left[a + \frac{b}{\left(\frac{E}{E_0\eta} + 1\right)}\right]\rho E + A\mu + B\mu^2, \text{ [if } \rho/\rho_0 > 1 \text{ or } (\rho/\rho_0 < 1 \;\&\; E < E_{\mathrm{iv}})] \quad \text{(S1)}$$

or

$$P(\rho, E) = a\rho E + \left[\frac{b\rho E}{\left(\frac{E}{E_0\eta}+1\right)} + A\mu e^{-\beta\left(\frac{\rho_0}{\rho} - 1\right)}\right] e^{-\alpha\left(\frac{\rho_0}{\rho} - 1\right)^2}, \text{ [if } (\rho/\rho_0 < 1 \;\&\; E > E_{\mathrm{cv}})] \quad \text{(S2)}$$

where $\eta = \rho/\rho_0$, $\mu = \eta - 1$, $\rho_0$ is the density under the reference state, *a*, *b*, $E_0$, *A*, *B*, *α*, *β* are the Tillotson parameters, and $E_{\mathrm{iv}}$ and $E_{\mathrm{cv}}$ are the internal energies required for incipient and complete vaporization, respectively. The equation (S1) is used in the compressed state ($\rho/\rho_0 > 1$) and for cold expanded states ($\rho/\rho_0 < 1$) where $E < E_{\mathrm{iv}}$, and the Eq. (S2) is applied to the hot expanded state ($\rho/\rho_0 < 1$ and $E > E_{\mathrm{cv}}$).

We determined the Tillotson parameters as follows. The obtained parameters and related thermos-elastic properties of Nylon 66 are summarized in Tables S2 and S3. First, we set the Tillotson parameters *a*, *α*, and *β* to be 0.5, 5, and 5, respectively, because these values are conventionally used [e.g., Melosh, 1989]. Second, we determined the constants *A* and *b* with the published data for Nylon 66. The elastic properties, including the reference density $\rho_0$ and the P– and S– wave speeds $V_{\mathrm{P}}$ and $V_{\mathrm{S}}$ for Nylon 66 can be taken from Carter and Marsh (1977). The bulk sound speed $C_{\mathrm{B}}$ is calculated as

$$C_{\mathrm{B}} = \sqrt{V_{\mathrm{P}}^2 - \frac{4}{3}V_{\mathrm{S}}^2}. \quad (3)$$

Since the Tillotson parameter *A* is equal to the isentropic bulk modulus, *A* is determined to be $A = \rho_0 C_{\mathrm{B}}^2$. The constant *b* is related to the Grüneisen constant under the reference state $\Gamma_0$ and the constant $a$ as $b = \Gamma_0 - a$. We employed the experimental data by Warfield et al. (1983) for $\Gamma_0$. Then, the constants $E_0$ and $B$ were determined by fitting the shock Hugoniot data by Carter and Marsh (1977). Figure S1 shows the wave propagation speed $U_S$ in Nylon 66 as a function of particle velocity $U_p$. We plotted the calculated $U_S$ values using the Tillotson EOS with our parameters along with the experimental data [Carter and Marsh, 1977].

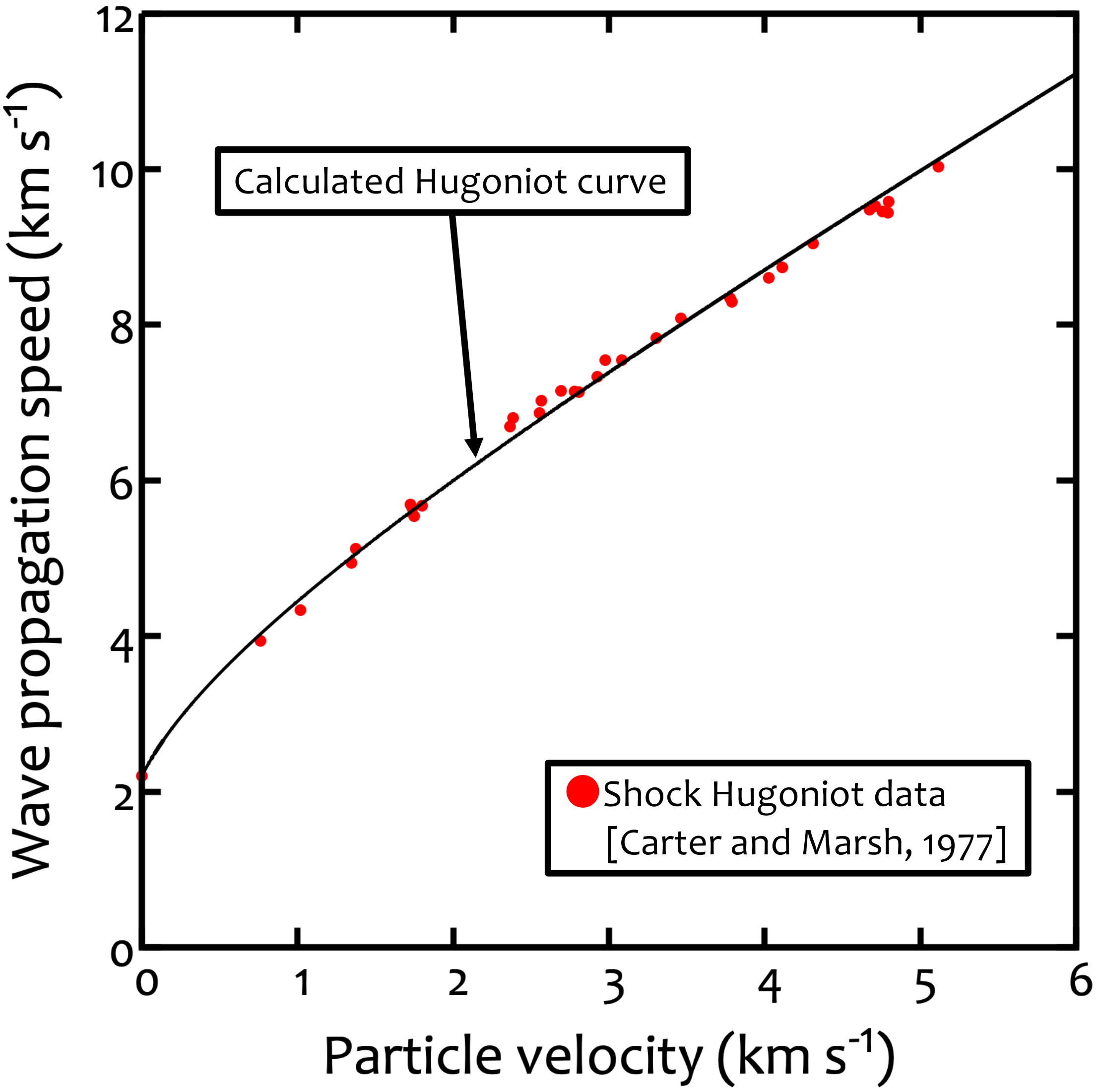

**Figure S1.** The Hugoniot curve of Nylon 66. Wave propagation speed is plotted as a function of the particle velocity behind the compressive wave front. The black solid line was obtained as the best-fit line with the Tillotson EOS. The parameters used in this calculation are listed in Table S2. The experimental data obtained by Carter and Marsh (1977) are also plotted as the filled red circles.

Next, we discuss the thermal properties, which are $C_P$, $E_{iv}$ and $E_{cv}$, where $C_P$ is isobaric specific heat. Since the vaporization behavior of polymers is rather complex, $E_{iv}$ and $E_{cv}$ cannot be tightly constrained. Here, we only provide first-order estimates as follows. Nylon 66 initiates thermal decomposition at ~700 K [e.g., Singh et al., 2003]. We can estimate $E_{iv}$ as

$$E_{iv} = \int_{T_o}^{T_d} C_p dT, \quad (4)$$

where $T_d$ is the temperature at thermal decomposition and $T_0$ is the initial temperature. The temperature-dependent $C_P$ pertaining to Nylon 66 was compiled by Gaur et al. (1983). Note that the $C_p$ data up to only 600 K was presented in Gaur et al. (1983). Since the Cp increase linearly against to temperature above the temperature of glass transition, we extrapolated the linear trend to $T_d$ (700 K). Figure S2 shows the isobaric specific heat and the enthalpy change as a function of temperature. Note that we added melting enthalpy $\Delta H_m$ at the melting point of Nylon 66. According to Fosse et al. (2019), $\Delta H_m$ pertaining to polymers are ranged from 0.1 to 0.3 MJ kg$^{-1}$. The constant $E_{cv}$ was estimated by following the procedure proposed by Sugita and Schultz (2003). The activation energy of the thermal decomposition of Nylon 66 was determined to be 0.36 MJ kg$^{-1}$ [Singh et al., 2003]. This value is the upper limit of the reaction heat of the endothermic thermal decomposition. Since the enthalpy change due to the phase change from solid to vapor for Nylon 66 was not reported in the literatures, we approximately used the value for olefins [Lide, 2001], which is typical hydrocarbons.

The iSALE only can treat a constant $C_p$ in the case with the Tillotson EOS although the $C_p$ of Nylon 66 depends on temperature as shown in Fig. S2. Then, the iSALE approximately calculated the temperature with the method by Ivanov et al. (2002) although the Tillotson EOS cannot explicitly provide temperature in the simulation. Here, we introduce the effective specific heat $C_{p,eff}$ of Nylon 66 as a constant. In shock physics codes, the melting temperature $T_m$ is particularly important because the yield strength of the shocked materials becomes zero above the melting point. Thus, we calculated $C_{p,eff}$ as

$$C_{p,eff} = \frac{\int_{T_0}^{T_m} C_p dT}{T_m - T_0}. \quad (5)$$

The required enthalpy to heat up from $T_0$ to $T_m$ is calculated to be $C_{p,eff}(T_m - T_0)$, which is the same in the case with the full treatment pertaining to $C_p(T)$.

**Table S2.** The Tillotson EOS parameters for Nylon 66.

| Variables | Values |
|---|---|
| Reference density (kg $m^{-3}$) | 1,146[a] |
| Isentropic bulk modulus $A$ (GPa) | 5.55[b] |
| Tillotson B constant (GPa) | 37[c] |
| Tillotson $E_0$ constant (MJ $kg^{-1}$) | 1.0[c] |
| Tillotson $a$ constant | 0.5[d] |
| Tillotson $b$ constant | 0.01[b] |
| Tillotson $\alpha$ constant | 5[d] |
| Tillotson $\beta$ constant | 5[d] |
| $E_{iv}$ (MJ $kg^{-1}$) | 1.2[b] |
| $E_{cv}$ (MJ $kg^{-1}$) | 2.0[b] |

[a]Carter and Marsh (1977)

[b]See text.

[c]Determined by fitting to the experimental data by Carter and Marsh (1977)

[d]Assumed by following Melosh (1989).

**Table S3.** Related thermo-elastic properties for Nylon 66

| | |
|---|---|
| P-wave speed (km $s^{-1}$) | 2.53[a] |
| S-wave speed (km $s^{-1}$) | 1.08[a] |
| Poisson's ratio | 0.389[b] |
| Compressive strength (MPa) | 100[c] |
| Molecular weight (kg $mol^{-1}$) | 0.22632[d] |
| Incipient decomposition temperature $T_d$ (K) | ~700[e] |
| Melting temperature (K) | 538[c] |
| Enthalpy change for melting (MJ kg $^{-1}$) | 0.1–0.3[f] |
| Activation energy for decomposition $E_a$ (MJ kg $^{-1}$) | 0.36[e] |
| Enthalpy change for vaporization $H_{vap}$ (MJ kg $^{-1}$) | ~0.4[g] |
| Reference Grüneisen constant $\Gamma_0$ | 0.49[h] |
| Effective specific heat (kJ $K^{-1}$ $kg^{-1}$) | 2.3[i] |

[a]Carter and Marsh (1977)

[b]Calculated using P- and S-wave speeds. The Poisson's ratio is required in shock physics modeling in the case with material strength.

[c]The values are taken from the CRC handbook edited by Haynes, (2010)

[d]Calculated based on the chemical formula of Nylon 66

[e]Singh et al. (2003)

[f]Fosse et al. (2019)

[g]Lide, (2001)

[h]Warfield et al. (1983)

[i]See text.

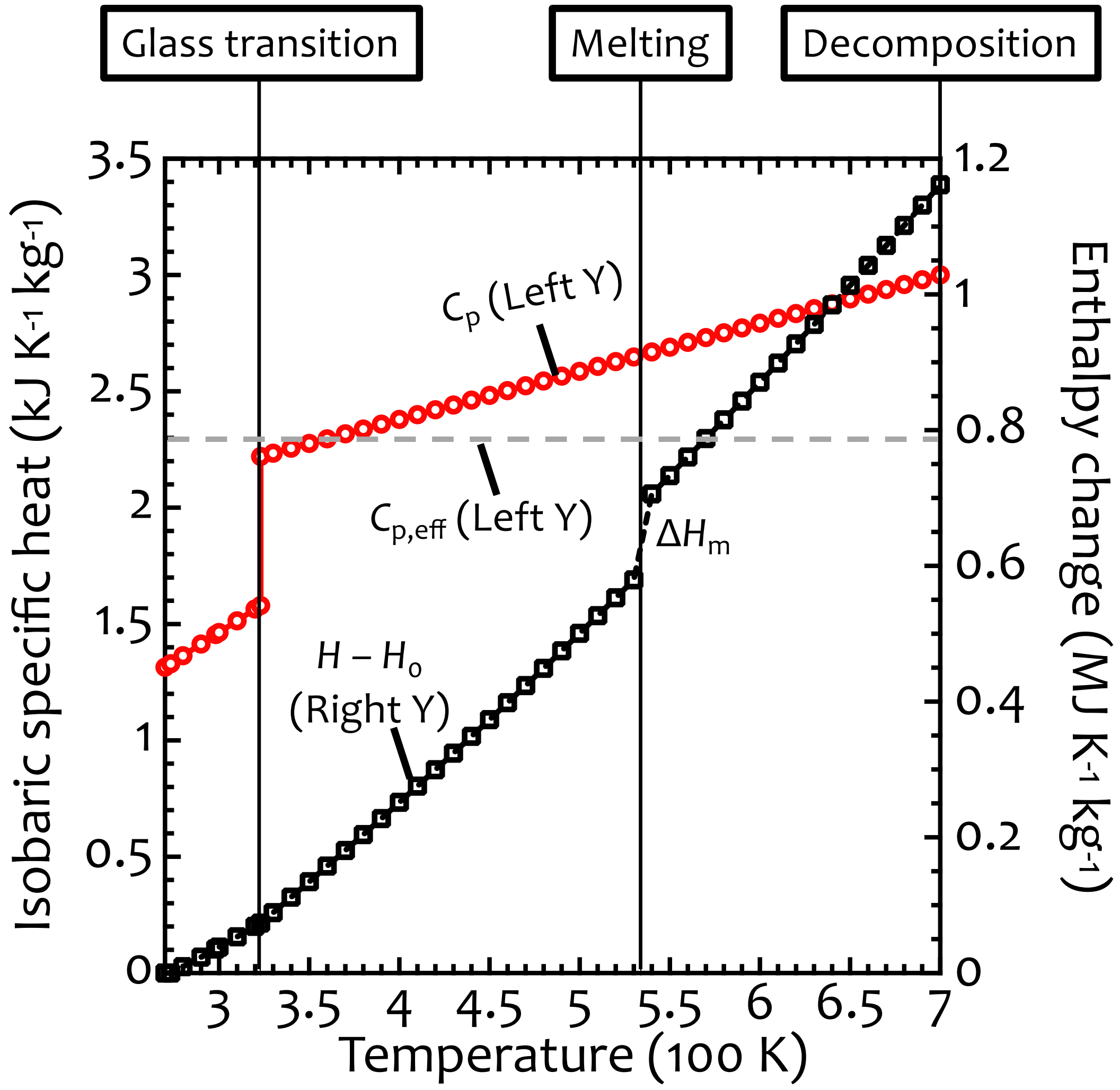

**Figure S2.** Isobaric specific heat $C_p$ (left Y) and the required enthalpy gain to heat up to a given temperature $H$ - $H_0$ (right Y) as a function of temperature. The temperature-dependent $C_p$ are taken from Gaur et al. (1983). The temperatures at glass transition, melting, and decomposition are shown in the figure. The effective specific heat $C_{p,eff}$ is also shown as the grey dashed horizontal line.